\documentclass{article}
\usepackage{spconf,amsmath,graphicx}

\usepackage{multirow}
\usepackage{graphicx}
\usepackage{amssymb, amsmath, bm, mathtools,array}
\usepackage{textcomp}
\usepackage{hyperref}
\usepackage{subcaption}
\usepackage{verbatim,lipsum}

\RequirePackage[normalem]{ulem} 
\RequirePackage{color}\definecolor{BLUE}{rgb}{0,0.20,0.75} 
\RequirePackage{color}\definecolor{BROWN}{RGB}{60,128,49} 

\newcommand{\bs}[1]{\boldsymbol{#1}}
\title{Effect of choice of probability distribution, randomness, and search methods for alignment modeling in sequence-to-sequence text-to-speech synthesis using hard alignment}
\name{Yusuke Yasuda$^{1,2}$, Xin Wang$^1$, Junichi Yamagishi$^{1,2}$}
\address{$^1$National Institute of Informatics, Japan
$^2$SOKENDAI, Japan\\
  {\small \tt yasuda@nii.ac.jp, wangxin@nii.ac.jp,  jyamagis@nii.ac.jp}}

\begin{document}
\ninept
\maketitle
\begin{abstract}

Sequence-to-sequence text-to-speech (TTS) is dominated by soft-attention-based methods. Recently, hard-attention-based methods have been proposed to prevent fatal alignment errors, 
but their sampling method of discrete alignment is poorly investigated. 
This research investigates various combinations of sampling methods and probability distributions for alignment transition modeling in a hard-alignment-based sequence-to-sequence TTS method called SSNT-TTS. We clarify the common sampling methods of discrete variables including greedy search, beam search, and random sampling from a Bernoulli distribution in a more general way. Furthermore, we introduce the binary Concrete distribution to model discrete variables more properly. The results of a listening test shows that deterministic search is more preferable than stochastic search, and the binary Concrete distribution is robust with stochastic search for natural alignment transition.

\end{abstract}
\begin{keywords}
speech synthesis, deep learning, sequence-to-sequence model
\end{keywords}

\section{Introduction}
\label{sec:intro}

Sequence-to-sequence TTS is a relatively new TTS approach. Because it can model complex alignments between source and target without an external aligner, a sequence-to-sequence TTS system can be trained in an end-to-end fashion, which makes it straightforward to leverage large-scale speech corpora. 
Among the various sequence-to-sequence frameworks, the one with attention is the most effective and has been used in many recently proposed TTS systems \cite{Wang2017, Sotelo2017Char2wavES, DBLP:conf/iclr/PingPGAKNRM18, Shen2017, DBLP:journals/corr/abs-1809-08895}. Some of these systems, e.g., Tacotron2 \cite{Shen2017}, can produce speech waveforms with human-level quality and naturalness. 
%
Despite the high-quality synthesized speech, attention-based sequence-to-sequence TTS systems suffer from alignment errors such as skipping and repetition, and some systems have even reported high alignment error rates \cite{DBLP:conf/iclr/PingPGAKNRM18, DBLP:journals/corr/abs-1905-09263, He2019}. Possible reasons for the alignment errors may include the undesirable flexibility of the soft attention used in those systems and the way of predicting alignment from soft attention.

Recently, we proposed a brand new sequence-to-sequence TTS framework based on hard attention \cite{Yasuda2019}. 
This framework was inspired by the segment-to-segment neural transducer (SSNT) \cite{yu2016online} and uses hard monotonic alignment instead of a soft one.
Thanks to the monotonic design, our SSNT-based TTS system, which is called SSNT-TTS, can avoid the fatal alignment errors that are commonly observed in soft-attention-based frameworks. However, we found that SSNT-TTS suffered different types of alignment errors, such as underestimation and overestimation of phone duration. As a result, SSNT-TTS did not outperform the major soft-attention-based methods such as Tacotron \cite{Wang2017}. 

SSNT-TTS represents hard alignment in probabilistic form and explicitly parameterizes the transition probability as a random variable. However, such an explicit way of modeling  alignment turns out to be challenging.
One issue is to how choose the proper probability distribution function (PDF) and a sampling method for alignment transition. 
In our previous work, we tried the simplest method based on greedy search \cite{Kato2019} and random sampling \cite{Yasuda2019}, and there remains obvious room for further improvements. In the present work, we investigate more sophisticated approaches for modeling the probability distribution and sampling alignment for SSNT-TTS. We also investigate the perceptual impact of the proposed approaches on the naturalness of synthetic speech. 

In Section 2 of this paper, we describe the background of SSNT-TTS, and in Section 3, we introduce the probability distribution and sampling methods for hard alignment. In Section 4, we discuss the experiments and results of the sampling methods under various conditions. We conclude in Section 5 with a brief summary and mention of future work..

\section{Background of SSNT-TTS}
SSNT-TTS models output acoustic features $\bs{y}_{1:J}$ given input linguistic features $\bs{x}_{1:I}$ (such as phonemes or text) by marginalizing the joint probability of $\bs{y}_{1:J}$ and alignment $\mathbf{z}$ conditioned on the input $\bs{x}_{1:I}$: 
\vspace{-5pt}
\begin{equation}
\label{eq:ssnt_eq1}
p(\bs{y}_{1:J} \mid \bs{x}_{1:I}) = \sum_{\forall \mathbf{z}} p(\bs{y}_{1:J}, \mathbf{z} \mid \bs{x}_{1:I}).
\vspace{-5pt}
\end{equation}
The joint probability in equation (\ref{eq:ssnt_eq1}) can be decomposed into products of alignment probability and output probability given the alignment at each time step under the first order Markov assumption:
\vspace{-5pt}
\begin{align}
\label{eq:factorized-joint-prob}
&p(\bs{y}_{1:J}, \mathbf{z} \mid \bs{x}_{1:I}) \approx \nonumber\\
&\prod_{j=1}^{J} p(z_j \mid z_{j-1}, \bs{y}_{1:j-1}, \bs{x}_{1:I}) p(\bs{y}_j \mid \bs{y}_{1:j-1}, z_j,  \bs{x}_{1:I}),
\vspace{-5pt}
\end{align}
where $z_j=i$ denotes the alignment between the input step $i\in\{1,\cdots,I\}$ and the output step $j\in\{1,\cdots,J\}$.

While the output probability in equation (\ref{eq:factorized-joint-prob}) can be an isotropic Gaussian distribution $p(\bs{y}_j \mid \bs{y}_{1:j-1}, z_j, \bs{x}_{1:I}) =\mathcal{N}(\bs{y}_j ; \boldsymbol{\mu}, \sigma^2\mathbf{I})$, the alignment probability can be represented by introducing binary alignment transition variable $a_{i,j} \in \{\mathtt{Emit}, \mathtt{Shift}\}$. This variable ensures the monotonic structure of the alignment: $\mathtt{Emit}$ means the alignment keeps the current input position (i.e., $\{z_{j-1}=i, z_{j}=i\}$), and $\mathtt{Shift}$ means the alignment proceeds to the next input position (i.e., $\{z_{j-1}=i-1, z_{j}=i\}$). Because there are only two ways for the alignment to reach $z_j = i$, namely, $\mathtt{Emit}$ transition from $z_{j-1} = i$ or $\mathtt{Shift}$ transition from $z_{j-1} = i - 1$, the alignment probability can be defined as \footnote{We simplified the alignment probability used in our previous work \cite{Yasuda2019} in which we defined 
$p(z_j = i \mid z_{j-1}, \bs{y}_{1:j-1}, \bs{x}_{1:I}) = p(a_{i-1,j} = \mathtt{Shift})p(a_{i,j} = \mathtt{Emit})$ if $z_{j-1} = i - 1$.}
\vspace{-5pt}
\begin{align}
\label{ssnt_tts_eq_transition_probability}
& p(z_j = i \mid z_{j-1}, \bs{y}_{1:j-1}, \bs{x}_{1:I}) = \nonumber\\
& \begin{cases}
p(a_{i,j} = \mathtt{Emit}) & z_{j-1} = i \\
p(a_{i-1,j} = \mathtt{Shift}) & z_{j-1}  = i - 1 \\
0 & z_{j-1}  < i - 1 \text{ or } z_{j-1} > i\\
\end{cases}
\vspace{-5pt}
\end{align}

The model can be optimized by minimizing the negative log likelihood $\mathcal{L}(\boldsymbol{\theta}) = -\log p(\bs{y}_{1:J} \mid \bs{x}_{1:I} ; \boldsymbol{\theta})$.
During prediction, output acoustic features are predicted given input $x_{z_j}$ at alignment $z_j$. The mean of the Gaussian distribution is used as the predicted output acoustic feature. The alignment is incremented at each time step as $z_{j} = z_{j-1} + k$, where $k = 0$ if $a_{z_{j-1},j} = \mathtt{Emit}$ and $k = 1$ if $a_{z_{j-1},j} = \mathtt{Shift}$. How to determine $k$ is the focus of this research. In previous work, $k$ was derived with greedy search \cite{Kato2019} or randomly sampled from a Bernoulli distribution \cite{Yasuda2019}. We generalize the sampling method of hard alignment in the next section.

\section{Sampling methods of hard alignment}
\label{sec:sampling-hard-aligments}

\subsection{Alignment sampling from a discrete distribution}
\label{subsec:sampling-from-discrete-dist}

Because SSNT-TTS is based on hard alignment, it not only predicts the probability values of alignment but also determines the best discrete alignment path during inference. This is quite different from soft attention, which predicts only probability values and uses them as the alignment to do a weighted sum over the encoder's output.
For the determination of the best discrete alignment path, SSNT-TTS samples alignment transition variable $a_{i,j} \in \{\mathtt{Emit}, \mathtt{Shift}\}$ from a learned probability distribution at each time step.

A well-known method to sample a random variable from a discrete distribution is the Gumbel-Max trick \cite{YELLOTT1977109}. The algorithm of the Gumbel-Max trick consists of two steps: one, sampling a continuous random variable from a learned distribution, and two, performing the argmax operation to output a final symbol. Let us denote the density of the binary transition variables as $p(a_{i,j} = \mathtt{Emit}) = \alpha_1$ and $p(a_{i,j} = \mathtt{Shift}) = \alpha_2$, where $\alpha_1 + \alpha_2 = 1$. To randomly sample a discrete alignment transition variable at input $i$ and time step $j$ with the Gumbel-Max trick, Gumbel noise is first added to each logit of the alignment transition densities, which can be written as
\begin{align}
\mathbb{P}(a_{i,j} = \mathtt{Emit}) &= \mathbb{P}(G_1 + \log\alpha_1 > G_2 + \log\alpha_2) \\
\label{gumbel-max-trick-for-bernouli}
                   &= \mathbb{P}(L + \log\alpha_1 >  \log\alpha_2),
\end{align}
where $G_1$ and $G_2$ are the Gumbel noises and $L = G_1 - G_2$. 
Since the difference $L$ is known to follow a Logistic distribution, a Logistic noise can be sampled by $L = \log(U) - \log(1 - U)$, where $U$ is a sample from the standard Uniform distribution.
%
%
Finally, the argmax operation is applied to obtain a discrete sample of the alignment transition variable
\begin{equation}
\label{eq:discrete-operation-in-sampling}
a_{i,j} = \mathrm{argmax}(L + \log\alpha_1, \log\alpha_2).
\end{equation}
%
This is a typical approach to draw random samples from the Bernoulli distribution, and we refer to this condition as “Logistic condition" because it involves sampling from the Logistic distribution in continuous space. We will introduce another continuous distribution for alignment transition distribution in Section \ref{subsec:continuous-relaxation}.

In implementation, because $\alpha_1 + \alpha_2 = 1$, we parameterize them as $\alpha_1 = \mathrm{sigmoid(\log\alpha, \lambda)}$ and $\alpha_2 = 1-\mathrm{sigmoid(\log\alpha, \lambda)}$, where $\text{sigmoid}(x,\lambda)=\frac{1}{1+\exp(-x/\lambda)}$ and $\lambda$ is a temperature hyper-parameter to control the shape of the sigmoid function (see Figure~\ref{fig:binary-concrete-distribution}). Accordingly, SSNT only needs to predict a single value of $\alpha$ at each time step. It can be shown that $\alpha = \alpha_1/\alpha_2$.

\subsection{Stochastic alignment search}
\label{subsec:stochastic-alignment-search}

Since SSNT-TTS treats the alignment as a latent variable, 
it must decode a single alignment path from all possible alignments during inference. From here on we refer to the process of selecting the best alignment as \textit{search}.

Greedy search is a search method that chooses a single path with a higher alignment transition probability at each time step, as
\vspace{-5pt}
\begin{equation}
\label{eq:greedy-decode}
a_{i,j} = \mathrm{argmax}(\log\alpha_1, \log\alpha_2)
= \begin{cases}
\mathtt{Emit} & \text{if } \alpha_1 > \alpha_2\\
\mathtt{Shift} & \text{if } \alpha_1 < \alpha_2
\end{cases}.
\end{equation}

From equations (\ref{eq:discrete-operation-in-sampling}) and (\ref{eq:greedy-decode}), we can see that the difference between the greedy search and random sampling from a Bernoulli distribution is just the Logistic noise $L$. 
For this reason, we refer to random sampling from a Bernoulli distribution as \textit{stochastic greedy search} for convenience.

A generalized version of greedy search is the beam search method.
The beam search method considers several candidates of alignments at each time step of decoding and selects the overall best candidate alignment path. The number of candidates is called beam width. Greedy search is a special case of beam search with beam width equal to one. Furthermore, by combining the sampling from the Bernoulli distribution and beam search, or more concretely, by adding the Logistic noise to the argmax operations during beam search, \textit{stochastic beam search} can be easily implemented.

\subsection{Continuous relaxation of discrete alignment variables}
\label{subsec:continuous-relaxation}


As we see in equation (\ref{gumbel-max-trick-for-bernouli}), the conventional random sampling from the Bernoulli distribution uses the Logistic distribution. The use of the Logistic distribution is, however, an oversimplified assumption since it has a symmetric shape and a single mode. 
More specifically, under this condition, similarity between the sample distribution and the actual discrete distribution mainly depends on the prediction accuracy of the $\alpha_1$ parameter. 

To model the hard alignment more appropriately, we propose using the \textit{binary Concrete distribution} \cite{DBLP:conf/iclr/MaddisonMT17} for the alignment transition probability. The Concrete distribution, which is also known as the Gumbel-Softmax distribution \cite{DBLP:conf/iclr/JangGP17}, is a continuous distribution that relaxes the discrete argmax operation. In this paper, we focus on the binary case of the Concrete distribution because our alignment transition variable $a_{i,j}$ has two states only.

\begin{figure}[t]
  \centering
  \includegraphics[width=\linewidth]{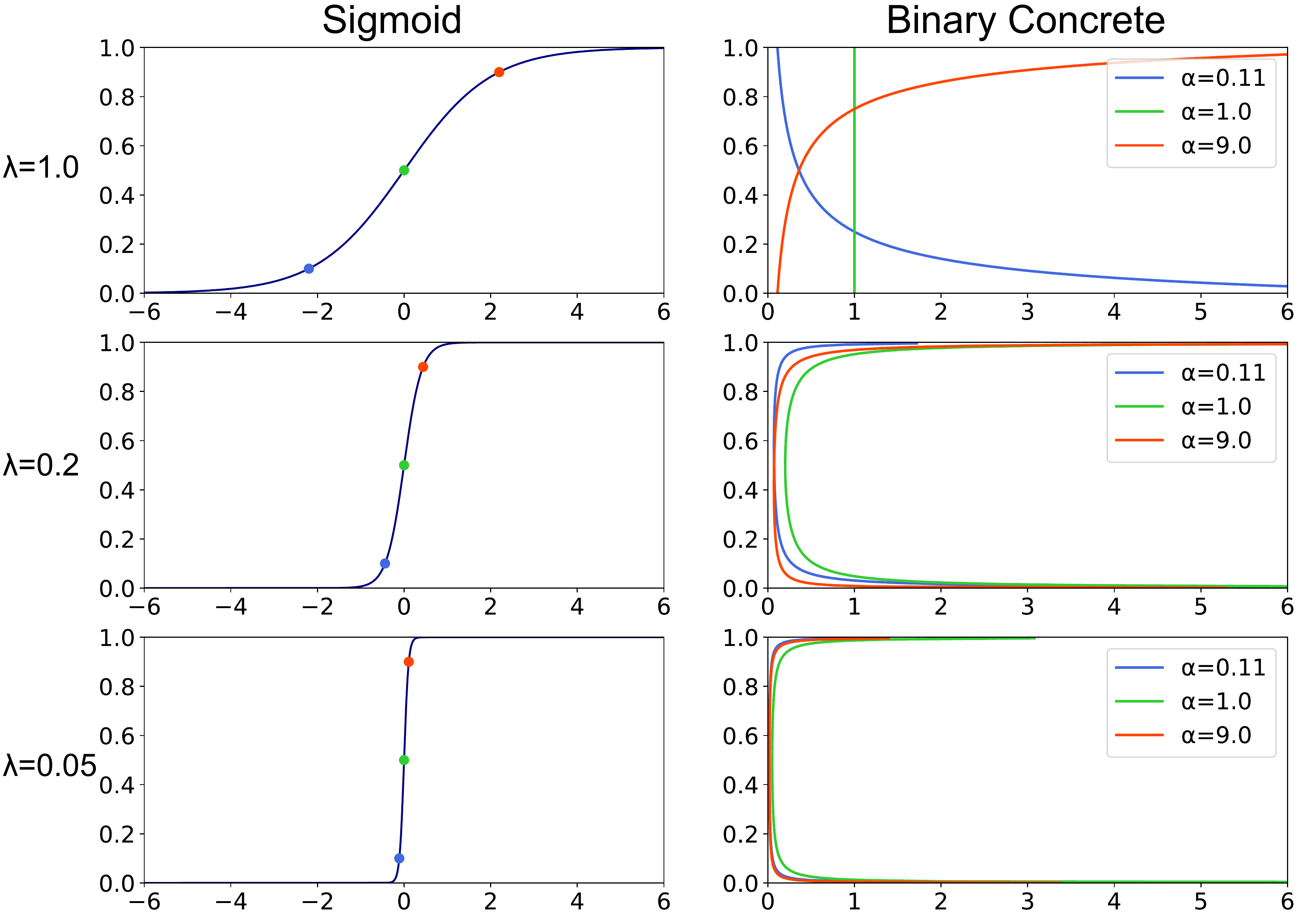}
  \vspace{-10pt}
  \caption{Left: Sigmoid function. Right: Binary Concrete distribution. Points on sigmoid function indicate $\alpha$ parameters for binary Concrete distribution.}
  \label{fig:binary-concrete-distribution}
  \vspace{-10pt}
\end{figure}

Using a concept similar to the Gumbel-Max trick, we can achieve random sampling from the binary Concrete distribution. Like the procedure described in Section \ref{subsec:sampling-from-discrete-dist}, random sampling from the binary Concrete distribution starts with adding the Logistic noise $L$ to a logit of the density of the alignment transition variable. Here, we introduce parameter $\alpha = \alpha_1/\alpha_2$, and consider the logit of the $\alpha$ instead. 
First, we add the Logistic noise $L$ to a sigmoid function:
\begin{equation}
\label{eq:sampling-from-binary-concrete-distribution}
\mathbb{P}(a_{i,j} = \mathtt{Emit}) = \frac{1}{1 + \exp\left(-(\log\alpha + L)/\lambda\right)}
\end{equation}
where $\lambda$ is a temperature parameter. This sigmoid function is used as a continuous version of the argmax function and a discrete sample can be obtained by applying the argmax operator to the output of the sigmoid function. The resulting distribution has the following form:
\begin{equation}
\label{eq:binconcrete-distribution}
\mathrm{BinConcrete}(x|\alpha, \lambda) = \frac{\lambda\alpha x^{-\lambda-1}(1 - x)^{-\lambda - 1}}{\left(\alpha x^{-\lambda} + (1 - x)^{-\lambda}\right)^2}.
\end{equation}
Figure \ref{fig:binary-concrete-distribution} shows the binary Concrete distribution with $\alpha$ and $\lambda$ of various values.
The parameter $\alpha$ can be learned by using the Logit noise during training. 
We refer to the condition using the binary Concrete distribution as the “binary Concrete condition".

\begin{figure}[t]
  \centering
  \includegraphics[width=\linewidth]{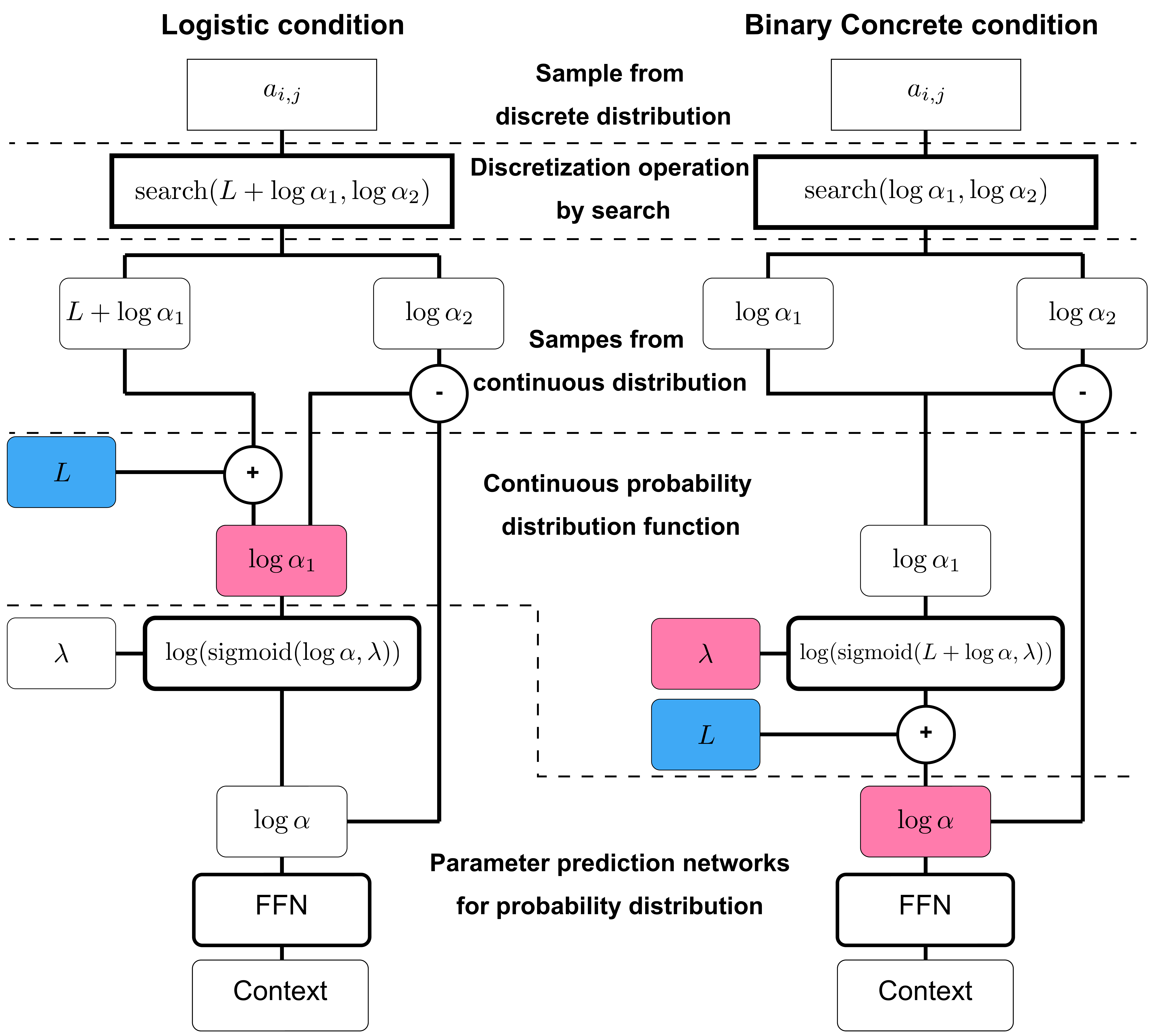}
  \vspace{-10pt}
  \caption{Sampling process of discrete alignment transition variable. Left: Logistic condition. Right: Binary Concrete condition. Bold boxes are functions, and normal boxes are values. Round boxes are continuous, and square boxes are discrete functions/values. Blue boxes are random noise, and red boxes are parameters of probability distribution. 
  }
  \label{fig:alignment-sampling-scheme}
  \vspace{-10pt}
\end{figure}

Figure \ref{fig:alignment-sampling-scheme} summarizes the sampling process of the Logistic and binary Concrete conditions implemented with neural networks. 
We can see that 
the both conditions have a similar network structure, and the difference is the location to add Logistic noise: the Logistic condition has the Logistic noise after the sigmoid function, while the binary Concrete condition has the Logistic noise before the sigmoid function.
Note that the use of Logistic noise is mandatory during training for the binary Concrete condition, as the randomness makes a model to learn the parameter $\alpha$ of the binary Concrete condition. In contrast, the Logistic condition does not use Logistic noise during training, since during training a model computes likelihood, not samples. At inference time, the use of the Logistic noise can be optional in both conditions. If the Logistic noise is omitted during inference, it is equivalent to deterministic search using mean value $\alpha_1$ for both the Logistic and binary Concrete conditions.

\section{Experiments}

\subsection{Experimental conditions}
We performed an experiment to compare the Logistic and binary Concrete conditions for SSNT-TTS. For each condition, we trained multiple SSNT-TTS models with different values of the  hyper-parameter $\lambda$, which can be 1.0, 0.2, or 0.05.
The effect of $\lambda$ on the Logistic and binary Concrete distributions is shown in Figure \ref{fig:binary-concrete-distribution}.

Given a trained SSNT-TTS model, we further compared combinations of different alignment search methods during inference. First, we compared the greedy and beam search methods. In the case of beam search, we used a beam width equal to 10.
For both greedy and beam search, we further compared deterministic and stochastic ways to determine the alignment transition action $a_{i,j}$ at each time step.
The deterministic way means absence of the Logistic noise while the stochastic way uses the Logistic noise during inference. 

\begin{figure}[t]
  \centering
  \includegraphics[width=0.8\linewidth]{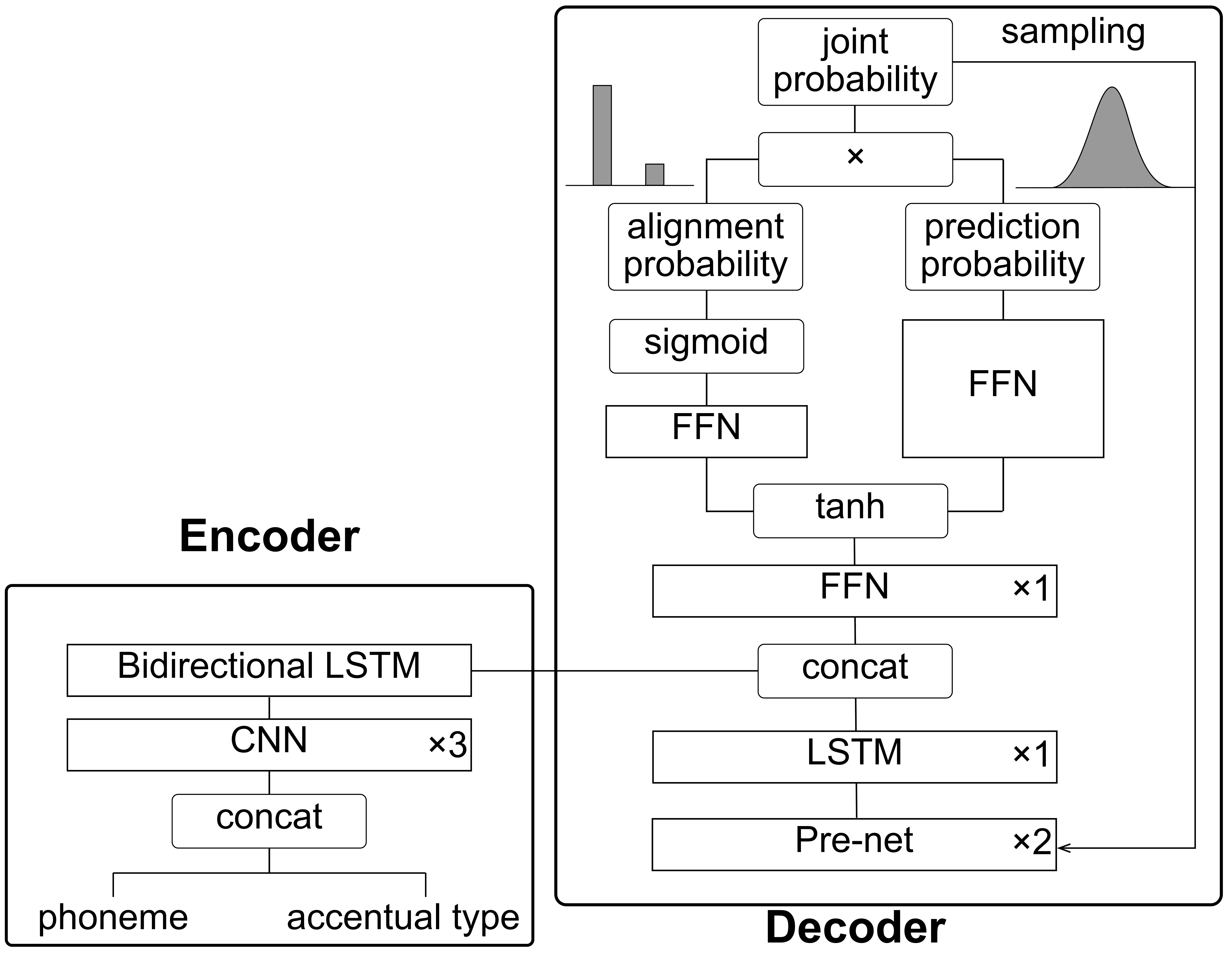}
  \vspace{-5pt}
  \caption{Network architecture of SSNT-TTS.}
  \label{fig:ssnt-architecture}
  \vspace{-10pt}
\end{figure}


We used the ATR Ximera dataset~\cite{kawai2006ximera}, a Japanese speech corpus containing 28,959 utterances totaling around 46.9 hours from a professional female speaker. We used manually annotated phoneme and accentual type labels~\cite{Luong2018}. The phoneme labels consist of 58 classes, including silences, pauses, and short pauses. All sentences start and end with a silence symbol. To train our proposed systems, we trimmed the beginning and ending silence from the utterances, after which the duration of the corpus was 33.5 hours. Fixed length silences were prepended and appended to target mel spectrograms. The frame size and shift used for the mel spectrogram were 5\,ms and 12.5\,ms, respectively.  We used 27,999 utterances for training, 480 for validation, and 480 for testing.

We used the same SSNT-TTS implementation as \cite{Yasuda2019}, except for the configuration of network parameters. Figure \ref{fig:ssnt-architecture} shows the network architecture of SSNT-TTS. Phoneme embedding vectors had 512 dimensions, and the accentual type embedding had 64 dimensions. For the encoder, we used the same conditions as \cite{Shen2017}. For the decoder, we used two pre-net \cite{Wang2017} layers with 256 and 128 dimensions, a single LSTM layer with 512 dimensions, and a single fully connected layer for context projection with 512 dimensions. We applied zoneout regularization~\cite{Krueger2016} to all LSTM layers with probability 0.1, as in \cite{Shen2017}. We used the Adam optimizer ~\cite{DBLP:journals/corr/KingmaB14} and trained the models with a batch size of 100 and a reduction factor equal to 2~\cite{Wang2017}. To reduce training time, we used mixed-precision training during which the computation was mainly done in half precision, but the model parameters were kept in normal precision. Furthermore, to prepare models in various conditions, we first created a seed model in the Logistic condition with temperature 0.05 running up to 270~K steps. Second, we derived models with temperature 1.0 and 0.2 by fine-tuning the seed model by 100~K steps. Finally, models with the BinConcrete condition were further trained by fine-tuning the models with the Logistic noise by 100~K steps. We used $\mu$-law WaveNet for the waveform generation~\cite{wavenet}\footnote{Audio samples can be found at our web page https://nii-yamagishilab.github.io/sample-ssnt-sampling-methods}.

\vspace{-10pt}
\subsection{Subjective evaluation}
\label{subsec:listening_test}

\begin{figure}[t]
  \centering
  \includegraphics[width=0.95\linewidth]{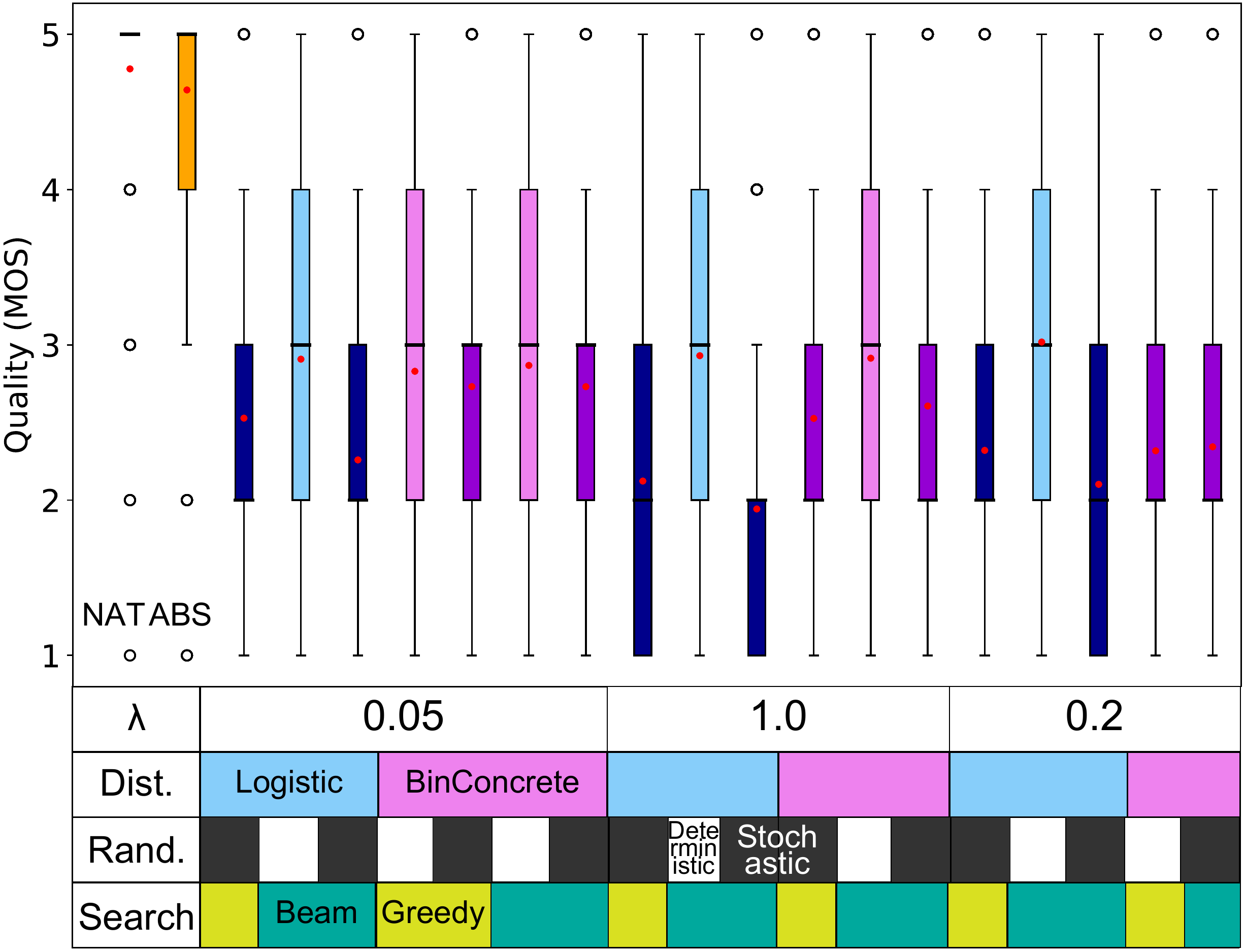}
  \caption{Results of listening test. Red circles indicate mean values. Bars show maximum, median, and minimum. NAT is natural sample, and ABS is analysis-by-synthesis.}
  \label{fig:mos}
  \vspace{-10pt}
\end{figure}

We organized a listening test to evaluate the proposed methods and recruited 193 native Japanese listeners by crowdsourcing. Listeners were presented with 40 samples from 20 systems and asked to judge the naturalness of speech using a five-point MOS score. The 20 systems included natural and analysis-by-synthesis samples in addition to the 18 SSNT-TTS models with different training and inference configurations, which are summarized at the bottom of Figure \ref{fig:mos}. One listener could evaluate at least three and at most ten sets. Each sample was evaluated three times, and we obtained 28,800 evaluations in total. We checked the statistical significance of MOS scores with a pairwise t-test.

Figure \ref{fig:mos} shows the results of the listening test. If we compare sampling conditions such as randomness and search method with the same temperature parameter $
\lambda$, we see a few interesting findings.

First, deterministic search conditions always outperformed stochastic search conditions.

Second, the effect of the search methods on the MOS scores depended on randomness, whether using deterministic search or stochastic search. Under the deterministic condition, systems with beam search always performed better than those with greedy search. On the other hand, under the stochastic and Logistic conditions, beam search performed worse than greedy search. For the BinConcrete condition, neither of the two search methods showed much difference.

If we focus on probability distribution variations while fixing temperature, we see that their scores depended on the use of randomness as well. With deterministic search, the Logistic conditions had slightly higher scores than the BinConcrete conditions, but their difference is not statistically significant. In contrast, stochastic search in the Logistic condition performed much worse than in the BinConcrete condition. The poor performance of stochastic search in the Logistic condition was mitigated by decreasing the temperature parameter.

Because the Logistic and BinConcrete conditions had similar scores under deterministic search, we can infer that both conditions were capable of estimating the alignment transition boundary. 
However, we also see that the Logistic condition does not parameterize a proper alignment transition distribution because it performed very badly under stochastic search, especially in 
stochastic beam search. 
Meanwhile, the BinConcrete conditions were relatively robust to stochastic search condition, presumably because 
it can fill the gap between continuous and discrete distributions.

\vspace{-5pt}
\section{Conclusion}
\label{sec:con}
\vspace{-5pt}

This paper investigated various combinations of sampling methods and probability distribution forms for hard-alignment transition modeling in SSNT-TTS. 
We have provided a more general view to clarify common sampling methods such as greedy search, beam search, and random sampling from a Bernoulli distribution by a combination of randomness and search methods.
In addition to sampling methods, we introduced the binary Concrete distribution to improve discrete alignment transition modeling by relaxing the argmax operation. Therefore, we enabled arbitrary combinations of probability distribution function, temperature parameter for probability distribution, the use of randomness for sampling, and search methods. We assessed the impact of various combinations of these parameters on the naturalness of synthetic speech in a listening test. Results showed that deterministic search was more favorable than stochastic search, and beam search improved naturalness. The binary Concrete distribution was relatively robust under stochastic search. In contrast, the Logistic distribution was not suitable for stochastic search.

Future work includes predicting the temperature $\lambda$ in a phoneme-dependent manner instead of fixing it as a hyper-parameter. 
Moreover, we are endeavoring to improve SSNT-TTS in general. The naturalness of our best SSNT-TTS system has a MOS score around 3, which means acceptable quality. It is still inferior to the current standard soft-attention-based TTS methods.

\vspace{0.5mm}
\noindent 
\textbf{Acknowledgements} 
This work was partially supported by a JST CREST Grant (JPMJCR18A6, VoicePersonae project), Japan, and by MEXT KAKENHI Grants (16H06302, 17H04687, 18H04120, 18H04112, and 18KT0051), Japan. The numerical calculations were carried out on the TSUBAME 3.0 supercomputer at the Tokyo Institute of Technology.

\bibliographystyle{IEEEtran}
\bibliography{main}

\begin{thebibliography}{10}
\providecommand{\url}[1]{#1}
\csname url@samestyle\endcsname
\providecommand{\newblock}{\relax}
\providecommand{\bibinfo}[2]{#2}
\providecommand{\BIBentrySTDinterwordspacing}{\spaceskip=0pt\relax}
\providecommand{\BIBentryALTinterwordstretchfactor}{4}
\providecommand{\BIBentryALTinterwordspacing}{\spaceskip=\fontdimen2\font plus
\BIBentryALTinterwordstretchfactor\fontdimen3\font minus
  \fontdimen4\font\relax}
\providecommand{\BIBforeignlanguage}[2]{{%
\expandafter\ifx\csname l@#1\endcsname\relax
\typeout{** WARNING: IEEEtran.bst: No hyphenation pattern has been}%
\typeout{** loaded for the language `#1'. Using the pattern for}%
\typeout{** the default language instead.}%
\else
\language=\csname l@#1\endcsname
\fi
#2}}
\providecommand{\BIBdecl}{\relax}
\BIBdecl

\bibitem{Wang2017}
Y.~Wang, R.~Skerry-Ryan, D.~Stanton, Y.~Wu, R.~J. Weiss, N.~Jaitly, Z.~Yang,
  Y.~Xiao, Z.~Chen, S.~Bengio, Q.~Le, Y.~Agiomyrgiannakis, R.~Clark, and R.~A.
  Saurous, ``Tacotron: Towards end-to-end speech synthesis,'' in \emph{Proc.
  Interspeech}, 2017, pp. 4006--4010.

\bibitem{Sotelo2017Char2wavES}
J.~Sotelo, S.~Mehri, K.~Kumar, J.~F. Santos, K.~Kastner, A.~Courville, and
  Y.~Bengio, ``Char2wav: End-to-end speech synthesis,'' in \emph{Proc. ICLR
  (Workshop Track)}, 2017.

\bibitem{DBLP:conf/iclr/PingPGAKNRM18}
W.~Ping, K.~Peng, A.~Gibiansky, S.~{\"{O}}. Arik, A.~Kannan, S.~Narang,
  J.~Raiman, and J.~Miller, ``{Deep Voice 3}: Scaling text-to-speech with
  convolutional sequence learning,'' in \emph{Proc. ICLR}, 2018.

\bibitem{Shen2017}
\BIBentryALTinterwordspacing
J.~Shen, R.~Pang, R.~J. Weiss, M.~Schuster, N.~Jaitly, Z.~Yang, Z.~Chen,
  Y.~Zhang, Y.~Wang, R.~Ryan, R.~A. Saurous, Y.~Agiomyrgiannakis, and Y.~Wu,
  ``Natural {TTS} synthesis by conditioning {WaveNet} on {Mel} spectrogram
  predictions,'' in \emph{Proc. ICASSP}, 2018, pp. 4779--4783. [Online].
  Available: \url{https://doi.org/10.1109/ICASSP.2018.8461368}
\BIBentrySTDinterwordspacing

\bibitem{DBLP:journals/corr/abs-1809-08895}
\BIBentryALTinterwordspacing
N.~Li, S.~Liu, Y.~Liu, S.~Zhao, M.~Liu, and M.~Zhou, ``Close to human quality
  {TTS} with transformer,'' \emph{CoRR}, vol. abs/1809.08895, 2018. [Online].
  Available: \url{http://arxiv.org/abs/1809.08895}
\BIBentrySTDinterwordspacing

\bibitem{DBLP:journals/corr/abs-1905-09263}
\BIBentryALTinterwordspacing
Y.~Ren, Y.~Ruan, X.~Tan, T.~Qin, S.~Zhao, Z.~Zhao, and T.~Liu, ``Fastspeech:
  Fast, robust and controllable text to speech,'' \emph{CoRR}, vol.
  abs/1905.09263, 2019. [Online]. Available:
  \url{http://arxiv.org/abs/1905.09263}
\BIBentrySTDinterwordspacing

\bibitem{He2019}
\BIBentryALTinterwordspacing
M.~He, Y.~Deng, and L.~He, ``Robust sequence-to-sequence acoustic modeling with
  stepwise monotonic attention for neural {TTS},'' in \emph{Proc. Interspeech
  2019}, 2019, pp. 1293--1297. [Online]. Available:
  \url{http://dx.doi.org/10.21437/Interspeech.2019-1972}
\BIBentrySTDinterwordspacing

\bibitem{Yasuda2019}
\BIBentryALTinterwordspacing
Y.~Yasuda, X.~Wang, and J.~Yamagishi, ``{Initial investigation of
  encoder-decoder end-to-end TTS framework using marginalization of monotonic
  hard alignments},'' in \emph{Proc. 10th ISCA Speech Synthesis Workshop},
  2019, pp. 211--216. [Online]. Available:
  \url{http://dx.doi.org/10.21437/SSW.2019-38}
\BIBentrySTDinterwordspacing

\bibitem{yu2016online}
L.~Yu, J.~Buys, and P.~Blunsom, ``Online segment to segment neural
  transduction,'' in \emph{Proceedings of the 2016 Conference on Empirical
  Methods in Natural Language Processing}, 2016, pp. 1307--1316.

\bibitem{Kato2019}
\BIBentryALTinterwordspacing
S.~Kato, Y.~Yasuda, X.~Wang, E.~Cooper, S.~Takaki, and J.~Yamagishi, ``{Rakugo
  speech synthesis using segment-to-segment neural transduction and style
  tokens — toward speech synthesis for entertaining audiences},'' in
  \emph{Proc. 10th ISCA Speech Synthesis Workshop}, 2019, pp. 111--116.
  [Online]. Available: \url{http://dx.doi.org/10.21437/SSW.2019-20}
\BIBentrySTDinterwordspacing

\bibitem{YELLOTT1977109}
\BIBentryALTinterwordspacing
J.~I. Yellott, ``The relationship between luce's choice axiom, thurstone's
  theory of comparative judgment, and the double exponential distribution,''
  \emph{Journal of Mathematical Psychology}, vol.~15, no.~2, pp. 109 -- 144,
  1977. [Online]. Available:
  \url{http://www.sciencedirect.com/science/article/pii/0022249677900268}
\BIBentrySTDinterwordspacing

\bibitem{DBLP:conf/iclr/MaddisonMT17}
\BIBentryALTinterwordspacing
C.~J. Maddison, A.~Mnih, and Y.~W. Teh, ``The concrete distribution: {A}
  continuous relaxation of discrete random variables,'' in \emph{Proc. ICLR
  (conference track)}.\hskip 1em plus 0.5em minus 0.4em\relax OpenReview.net,
  2017. [Online]. Available: \url{https://openreview.net/forum?id=S1jE5L5gl}
\BIBentrySTDinterwordspacing

\bibitem{DBLP:conf/iclr/JangGP17}
\BIBentryALTinterwordspacing
E.~Jang, S.~Gu, and B.~Poole, ``Categorical reparameterization with
  gumbel-softmax,'' in \emph{Proc ICLR (conference track)}.\hskip 1em plus
  0.5em minus 0.4em\relax OpenReview.net, 2017. [Online]. Available:
  \url{https://openreview.net/forum?id=rkE3y85ee}
\BIBentrySTDinterwordspacing

\bibitem{kawai2006ximera}
H.~Kawai, T.~Toda, J.~Yamagishi, T.~Hirai, J.~Ni, N.~Nishizawa, M.~Tsuzaki, and
  K.~Tokuda, ``Ximera: A concatenative speech synthesis system with large scale
  corpora,'' \emph{IEICE Transactions on Information and System (Japanese
  Edition)}, pp. 2688--2698, 2006.

\bibitem{Luong2018}
H.-T. Luong, X.~Wang, J.~Yamagishi, and N.~Nishizawa, ``Investigating accuracy
  of pitch-accent annotations in neural-network-based speech synthesis and
  denoising effects,'' in \emph{Proc. {Interspeech}}, 2018, pp. 37--41.

\bibitem{Krueger2016}
D.~Krueger, T.~Maharaj, J.~Kram{\'a}r, M.~Pezeshki, N.~Ballas, N.~R. Ke,
  A.~Goyal, Y.~Bengio, A.~Courville, and C.~Pal, ``Zoneout: Regularizing rnns
  by randomly preserving hidden activations,'' in \emph{Proc. ICLR}, 2017.

\bibitem{DBLP:journals/corr/KingmaB14}
D.~P. Kingma and J.~Ba, ``Adam: A method for stochastic optimization,'' in
  \emph{Proc. ICLR}, 2014.

\bibitem{wavenet}
\BIBentryALTinterwordspacing
A.~van~den Oord, S.~Dieleman, H.~Zen, K.~Simonyan, O.~Vinyals, A.~Graves,
  N.~Kalchbrenner, A.~W. Senior, and K.~Kavukcuoglu, ``Wavenet: {A} generative
  model for raw audio,'' \emph{CoRR}, vol. abs/1609.03499, 2016. [Online].
  Available: \url{http://arxiv.org/abs/1609.03499}
\BIBentrySTDinterwordspacing

\end{thebibliography}

\end{document}